\documentclass{aastex63}

\usepackage{color} 
\usepackage{amsmath}
\usepackage{bm}
\usepackage{mathtools}

\newcommand{\tred}[1]{#1}
\newcommand{\figdir}{.}

\begin{document}

\title{
  A new broadening technique of numerically unresolved solar transition region \\
  and its effect on the spectroscopic synthesis using coronal approximation
}

\correspondingauthor{Haruhisa Iijima}
\email{h.iijima@isee.nagoya-u.ac.jp}

\author{Haruhisa Iijima}
\affiliation{
  Institute for Space-Earth Environmental Research, Nagoya University, Furocho, Chikusa-ku, Nagoya, Aichi 464-8601, Japan
}
\affiliation{
  National Astronomical Observatory of Japan, 2-21-1 Osawa, Mitaka, Tokyo 181-8588, Japan
}

\author{Shinsuke Imada}
\affiliation{
  Institute for Space-Earth Environmental Research, Nagoya University, Furocho, Chikusa-ku, Nagoya, Aichi 464-8601, Japan
}

\begin{abstract}
The transition region is a thin layer of the solar atmosphere that controls the energy loss from the solar corona.
Large numbers of grid points are required to resolve this thin transition region fully in numerical modeling.
In this study, we propose a new numerical treatment, called LTRAC, which can be easily extended to the multi-dimensional domains.
We have tested the proposed method using a one-dimensional hydrodynamic model of a coronal loop in an active region.
The LTRAC method enables modeling of the transition region with the numerical grid size of 50--100 km, which is about 1000 times larger than the physically required value.
We used the velocity differential emission measure to evaluate the possible effects on the optically thin emission.
Lower temperature emissions were better reproduced by the LTRAC method than by previous methods.
Doppler shift and non-thermal width of the synthesized line emission agree with those from a high-resolution reference simulation within an error of several km/s above the formation temperature of $10^5$ K.
\end{abstract}

\keywords{
  hydrodynamics --- magnetohydrodynamics (MHD) --- Sun: transition region
  --- Sun: chromosphere --- Sun: corona --- Sun: flares
}

\section{Introduction}
\label{sec:intro}

The solar transition region is a thin layer between the chromosphere and the corona.
In this region, the temperature and mass density change steeply from chromospheric values to coronal values.
The rapid change of the mass density produces the steep gradients of the acoustic and Alfv\'{e}n speeds, which leads to wave reflection.
The rate of reflection of Alfv\'{e}n waves at the thin transition region is roughly determined by the density variation in that region \citep{1984SoPh...91..269H,2012A&A...538A..70V}.
The free magnetic energy injected into the solar corona is released through impulsive heating events in both the magnetic braiding and the wave heating models \citep[e.g.,][]{2011ApJ...736....3V,2017ApJ...834...10R,2018MNRAS.476.3328M}.
The thermal energy in the solar corona is transported by thermal conduction down to the transition region, releasing the energy as radiation into space.
The resulting evaporation of the chromospheric plasma increases the coronal density \citep{2010LRSP....7....5R,2015RSPTA.37340256K}.
Thus, accurate modeling of the transition region cannot be avoided in modeling the energy exchange between the chromosphere and the corona.

A simple static, coronal-loop model was introduced by \cite{1978ApJ...220..643R}.
The major assumptions in this model are:
(1) a coronal loop is sufficiently short so that the spatial variation of the gas pressure can be ignored,
(2) a volumetric heating rate in a coronal loop is spatially uniform,
and (3) the energy balance in a static loop is given by
\begin{equation}
  \frac{{\partial}}{{\partial}{s}}\left(
    {\kappa}\frac{{\partial}{T}}{{\partial}{s}}
  \right)+Q_\mathrm{ext}+Q_\mathrm{rad}=0,
\end{equation}
where $T$ is the temperature, $s$ is the coordinate along the coronal loop, $\kappa$ is the thermal conduction coefficient, $Q_\mathrm{ext}$ is the volumetric heating rate, and $Q_\mathrm{rad}$ is the optically thin radiative cooling rate.
In this model, the temperature profile near the transition region is sensitive to the gas pressure (or heating rate) but is not sensitive to the length of the loop.
\tred{
We estimated the thickness of the transition region using the spatial profile of the temperature given by Eq. (C1) in Appendix C of \cite{1978ApJ...220..643R}.
The thickness of the transition region was defined as a distance $s(T_1)-s(T_0)$, where $T_0=2\times10^4$ K (i.e., the temperature at the footpoint of the coronal loop), $T_1=10^5$ K (the typical temperature at the transition region), and $s(T)$ is the coordinate $s$ at the temperature $T$.
The estimated thickness of the transition region was approximately $s(T_1)-s(T_0){\sim}28\times(T_\mathrm{top}\text{ [K]}/10^6)^{-3}$ km, where $T_\mathrm{top}$ is the temperature at the apex of the coronal loop.
}
Even the quiet corona cannot be resolved by a grid size of several tens of kilometers, which is typically assumed in multidimensional simulations \citep[e.g.][]{2011A&A...531A.154G,2015ApJ...812L..30I,2017ApJ...848...38I,2017ApJ...834...10R,2007ApJ...665.1469A}.
The transition region becomes even thinner in hotter active regions or flaring loops due to the strong temperature sensitivity of the thermal conduction coefficient and the radiative cooling.

The lack of sufficient resolution of the transition region in hydrodynamic and magnetohydrodynamic simulations has been a topic of continuing discussion \citep{1982SoPh...76..331C,2006SoPh..234...41K,2013ApJ...770...12B}.
One possible solution is to use a moving mesh \citep{1992ApJ...397L..59C} or adaptive mesh refinement \citep{2011ApJS..194...26B}.
However, it is difficult to achieve such small numerical grid sizes in multidimensional simulations due to the resulting high computational cost.
As the coronal heating processes (both magnetic braiding and wave heating) are intrinsically three-dimensional in nature, it is necessary to devise a way to capture the transition region adequately on a coarse numerical grid to model the solar corona realistically.

A technique for broadening the numerical transition region has been suggested by \cite{2001JGR...10625165L}, \cite{2009ApJ...690..902L}, and \cite{2013ApJ...773...94M}.
By introducing the problem-dependent parameter $T_c$, they suggested that the thermal conduction coefficient should be enhanced by a factor $f_c=\max(1,T/T_c)$ to broaden the unresolved transition region.
They also reduced the radiative cooling rate by the same factor $f_c$ so that the total radiative loss remains unchanged.
\cite{2019ApJ...873L..22J} and \cite{2020A&A...635A.168J} suggested a way to model the transition using a method they term Transition Region Adaptive Conduction (TRAC).
This method automatically determines the optimal value of the parameter $T_c$ in one-dimensional coronal loop simulations.
By selecting the highest temperature at a location with a high temperature gradient, they showed that the dynamic evolution of a coronal loop can be reproduced with the coarse grid size of 100 km.
However, a multidimensional extension of the TRAC method is not straightforward, because we need a way to track the magnetic field line as the optimal value of $T_c$ is determined along the field line {\citep{2021A&A...648A..29Z}}.
\cite{2017A&A...597A..81J,2017A&A...605A...8J} also suggested an alternative way to resolve the transition region using jump conditions.

Spectroscopic observations in the EUV and X-ray are important diagnostic tools for the solar corona and transition region, especially for information about plasma motions \citep[e.g.,][]{1998ApJS..114..151C,1999A&A...349..636T,2007ApJ...667L.109D,2008ApJ...679L.155I}.
In particular, the statistical characteristics of non-thermal line widths and the Doppler velocities have been used as constraints on coronal heating models \citep[e.g.,][]{1999ApJ...522.1148P,2009ApJ...705L.208I,2016ApJ...820...63B,2016ApJ...827...99T,2017ApJ...849...46V}.
As the line profiles contain considerable information about the coronal plasma, a numerical treatment of the broadening method must therefore reproduce the spectroscopic observables.

In this study, we propose a new approach called the LTRAC (localized TRAC) method to broaden unresolved transition region in numerical simulations.
The local nature of the LTRAC method allows us to apply it to the parallel computation of multidimensional simulations using the domain decomposition technique.
As a large number of grid points is required to resolve the transition region properly, we tested the proposed method using a one-dimensional hydrodynamic model of a solar coronal loop.
We investigated the performance of the proposed method by examining the time evolution of the plasma in the transition region and corona, specifically focusing on spectroscopic observables in the optically thin approximation.
We found that the LTRAC method reproduces the spectroscopic observables coming from the low temperature plasma better than previous methods.

\section{Method}
\label{sec:method}

\subsection{Simulation setup}
\label{subsec:num_setup}

To investigate the performance of the methods used to broaden the transition region, we solved the one-dimensional hydrodynamic equations along a field line, including the effects of gravity, Spitzer-H\"{a}rm conduction, radiative cooling, and external heating.
The basic equations are the continuity equation
\begin{equation}
  \frac{{\partial}{\rho}}{{\partial}{t}}
  +\frac{{\partial}}{{\partial}{s}}\left({\rho}{V_s}\right)
  =0,
\end{equation}
the equation of motion
\begin{equation}
  \frac{{\partial}}{{\partial}{t}}\left({\rho}{V_s}\right)
  +\frac{{\partial}}{{\partial}{s}}\left({\rho}{V_s^2}\right)
  +\frac{{\partial}{P}}{{\partial}{s}}
  ={\rho}{g_s},
\end{equation}
and the internal energy equation
\begin{equation}
  \frac{{\partial}{e}}{{\partial}{t}}
  +\frac{{\partial}}{{\partial}{s}}\left[\left(e+P\right){V_x}\right]
  =V_s\frac{{\partial}{P}}{{\partial}{s}}
  +Q_{\rm cnd}+Q_{\rm rad}+Q_{\rm ext},
\end{equation}
where $\rho$ is the mass density, $P$ is the gas pressure, $e$ is the internal energy density, {$V_s$} is the velocity along a field line, and $g_s$ is the gravitational acceleration along a field line.
The equation of state is computed assuming local thermodynamic equilibrium and considering the most abundant six elements in the solar atmosphere.
The ionization energies of the six elements and the latent internal energy of molecular hydrogen are included in the internal energy of the plasma.

The quantity $Q_\mathrm{cnd}$ represents the volumetric heating rate due to the thermal conduction:
\begin{equation}
  \label{eq:qcnd}
  Q_{\rm cnd}=
  \frac{\partial}{{\partial}{s}}
  \left(
    \kappa\frac{{\partial}{T}}{\partial{s}}
  \right).
\end{equation}
For the thermal conduction coefficient, we assume Spitzer-H\"{a}rm conduction $\kappa={\kappa_0}T^{5/2}$ with $\kappa_0=10^{-6}$ in c.g.s. units.

The radiative cooling rate $Q_\mathrm{rad}$ is a combination of the cooling rates from the optically thick and thin regions, as given by
\begin{equation}
  \label{eq:qrad}
  Q_{\rm rad}=\left(1-\xi\right)Q_{\rm thick}+{\xi}Q_{\rm thin},
\end{equation}
where $\xi=\exp\left(-{P}/{P_{\rm thick}}\right)$ is the switching function between optically thick cooling, $Q_\mathrm{thick}$, and thin cooling, $Q_\mathrm{thin}$.
We chose $P_\mathrm{thick}=10^4$ dyn/cm$^2$ as the threshold parameter.
The optically thick cooling rate is given by
\begin{equation}
  Q_{\rm thick}=-{\rho}C_V\frac{T-T_{\rm rad}}{\tau},
\end{equation}
where $C_V$ is the isochoric heat capacity per unit mass, and $T_\mathrm{rad}=6000$ K is the radiation temperature.
Following the approach of \cite{2005ApJ...618.1020G}, we chose the cooling timescale $\tau$ to be
\begin{equation}
  \tau=\max\left(0.1\sqrt{\frac{\rho_{\rm surf}}{\rho}},4\Delta{t}\right),
\end{equation}
where $\rho_\mathrm{surf}=2\times10^{-7}$ g/cm$^3$.
We used the slightly modified version of the optically thin cooling rate given by
\begin{equation}
  Q_{\rm thin}=-n_en_H\Lambda(T)
  \left[1-\left(\frac{T_{\rm rad}}{T}\right)^4\right]
\end{equation}
to prevent the formation of low temperature regions with $T<T_\mathrm{rad}$.
Here, $n_e$ and $n_H$ represent the number densities of electrons and hydrogen nuclei.
This modification does not affect either the transition region or the corona.

The numerical scheme is based on the finite difference method.
The hyperbolic equations are solved using the second-order SLIP scheme \citep{1995IJCFD...4..171J} with the third-order strongly stability preserving Runge-Kutta method \citep{1988JCoPh..77..439S}.
The energy-consistent formulation by \cite{2021JCoPh.43510232I} is used in the spatial discretization to ensure the total energy conservation while explicitly solving the internal energy equation.
We used the operator splitting between the hydrodynamic equations and the thermal conduction equation.
The thermal conduction equation is discretized using the second-order central difference method, and it is integrated implicitly using the backward Euler method to avoid the severe time step restriction encountered in explicit schemes.

The numerical domain extends from the solar surface ($s=0$) to the top of the coronal loop ($s=L$).
The half length of the coronal loop $L$ is chosen to be 25 Mm in a typical case.
We assumed a semi-circular coronal loop and reduced the field-aligned gravitational acceleration $g_s$ following the geometrical effect.
We calculated the initial temperature profile from
\begin{equation}
  \ln{T(s)}=\ln{T_{\rm surf}}
  +\frac{1}{2}\left[1+\tanh\left(\frac{s-s_{\rm TR}}{\Delta{s}_{\rm TR}}\right)\right]
  \left(\ln{T_{\rm surf}}-\ln{T_{\rm top}}\right),
\end{equation}
where $s_\mathrm{TR}=2.5$ Mm, ${\Delta}{s}_\mathrm{TR}=500$ km, $T_\mathrm{surf}=6000$ K, and $T_\mathrm{top}=10^6$ K.
The mass density is integrated assuming hydrostatic equilibrium, with the boundary value of $\rho=2\times10^{-7}$ g/cm$^3$.
This initial condition is not energetically balanced because of the thermal conduction, radiative cooling, and external heating terms.
We assume reflective boundary conditions at the top and bottom of the domain.
In this study, the dynamical evolution of the simulated corona is controlled by imposing an external heating rate $Q_\mathrm{ext}$ to mimic coronal heating due to arbitrary processes.
In a typical case, we impose spatially uniform volumetric heating rate $Q_\mathrm{ext}=F_\mathrm{ext}/L$.
The injected energy flux $F_\mathrm{ext}$ is set to $3\times10^7$ erg/cm$^2$/s for the initial $600$ s of the simulation and set to zero after the time of 600 s.
This setup mimics a coronal loop in an active region.

\subsection{The LTRAC method}
\label{subsec:ltrac}

We now describe the LTRAC method we implemented to broaden the unresolved transition region in the model described in Sec. \ref{subsec:num_setup}.
As our new method is an extension of the methods suggested by \cite{2009ApJ...690..902L} and \cite{2019ApJ...873L..22J}, we summarize their methods first.

\cite{2009ApJ...690..902L} used energy balance between the radiative cooling and conductive heating,
\begin{equation}
  Q_{\rm rad}{\sim}\frac{{\kappa}T}{\delta_\mathrm{TR}^2},
\end{equation}
to obtain the approximate thickness of the transition region as:
\begin{equation}
  \label{eq:delta_tr}
  \delta_\mathrm{TR}{\sim}\sqrt{\frac{\kappa{T}}{Q_{\rm rad}}}.
\end{equation}
\tred{
Here, we note that, if we assume that the typical temperature in the transition region does not depend on the temperature at the loop top, the temperature dependence of the transition region thickness (discussed in Sec. \ref{sec:intro}) can be derived as $\delta_\mathrm{TR}\propto{T_\mathrm{top}^{-3}}$, where we used the scaling law of Eq. (4.3) in \cite{1978ApJ...220..643R}.
The dependence on the loop top temperature $T_\mathrm{top}$ comes from the dependence on the mass density of the radiative cooling rate $Q_\mathrm{rad}$.
}

\tred{
\cite{2009ApJ...690..902L} proposed that, if $\kappa$ and $Q_\mathrm{rad}$ are substituted by $f_c\kappa$ and $Q_\mathrm{rad}/f_c$, respectively, the transition region can be broadened by a factor of $f_c$.
This substitution can be alternatively expressed by
\begin{equation}
  \label{eq:limit_kappa_qrad}
  \kappa^\mathrm{b}=f_c\kappa
  \text{ and }
  Q_\mathrm{rad}^\mathrm{b}=Q_\mathrm{rad}/f_c,
\end{equation}
where the variable with superscript $U^\mathrm{b}$ indicates the variable $U$ modified by the broadening technique of the transition region.
The substitution (\ref{eq:limit_kappa_qrad}) broadens the approximate thickness of the transition region (\ref{eq:delta_tr}) as $\delta_\mathrm{TR}^\mathrm{b}=f_c\delta_\mathrm{TR}$.
}
We did not include the effect of the heat flux saturation \citep{1985ApJ...289..414F,2005ApJ...628.1023P,2006A&A...458..987B} for simplicity, which is important for very hot coronal loops.
\tred{
One possible way to implement the heat flux saturation is to enhance the saturated (limited) heat flux (or conductive flux) by a factor of $f_c$ instead of the thermal conduction coefficient $\kappa$.
However, its validity and performance should be investigated in the future study.
}

\cite{2020A&A...635A.168J} pointed out that the substitution (\ref{eq:limit_kappa_qrad}) does not alter the total radiative loss from the broadened transition region.
From Eq. (\ref{eq:delta_tr}), the total radiative loss integrated across the transition region can be approximated as
\begin{equation}
  \label{eq:net_qrad_unchange}
  \int_{s_0}^{s_1}Q_{\rm rad}ds
  {\sim}Q_{\rm rad}{\delta_\mathrm{TR}}
  {\sim}\sqrt{{\kappa}{Q_{\rm rad}}{T}}
  {=}\sqrt{{\kappa^\mathrm{b}}{Q_{\rm rad}^\mathrm{b}}{T}},
\end{equation}
where $s_0$ and $s_1$ are the spatial coordinates at the bottom and top of the transition region, respectively, \tred{and we used Eq. (\ref{eq:limit_kappa_qrad}) to derive the last equality.
}
As the substitution of $\kappa$ and $Q_\mathrm{rad}$ by Eq. (\ref{eq:limit_kappa_qrad}) does not alter the right-hand-side of Eq. (\ref{eq:net_qrad_unchange}), the total radiative loss from the transition region is not altered even if the transition region is numerically broadened.
Similarly, the total conductive heating of the transition region (${\kappa}{T}/\delta_\mathrm{TR}$) remains unchanged by the substitution (\ref{eq:limit_kappa_qrad}).

\cite{2009ApJ...690..902L} suggested to determine the broadening factor $f_c$ as
\begin{equation}
  f_c=\max(1,T_c/T)^{5/2},
\end{equation}
where $T_c$ is a parameter that depends on the specific problem and the grid size.
For example, $T_c$ is set to a typical temperature at the top of the transition region.
\cite{2009ApJ...690..902L} noted that the lower $T_c$ requires higher numerical resolution to resolve the transition region.
\cite{2019ApJ...873L..22J} proposed the TRAC method that determines the optimal value of $T_c$ from
\begin{equation}
  T_c=\max_s\left(T(s)\right)\text{ s.t. }
  \left|\frac{\Delta{s}}{T}\frac{{\partial}T}{{\partial}{s}}\right|>\sigma_c,
\end{equation}
where $\varDelta{s}$ is the grid size, and the problem independent parameter $\sigma_c=1/2$ is introduced.
Following \cite{2019ApJ...873L..22J}, the variable $T_c$ is limited by imposing a lower bound of typical chromospheric temperature ($2\times10^4$ K) and an upper limit of the 20 percents of the maximum coronal temperature.
Note that a multidimensional extension of the TRAC method is not straightforward, as the maximum is taken along a field line.
In parallel simulations with the domain decomposition, the computation of the maximum value along a field line requires global communication, which may result in less parallel efficiency.

There are slight differences between the implementations of the TRAC method in \cite{2019ApJ...873L..22J} and \cite{2020A&A...635A.168J}.
Our implementation of the TRAC method is based on \cite{2019ApJ...873L..22J}, not \cite{2020A&A...635A.168J}.
\cite{2020A&A...635A.168J} suggested to divide the external heating rate by the factor $f_c$.
We do not follow their approach because the modification of the external heating rate may cause an error in the total energy injected into the whole atmosphere.
As we assume a spatially uniform volumetric heating rate, this minor difference in the heating term does not affect the results presented in this paper.
\cite{2020A&A...635A.168J} also suggested to use the temporal smoothing on $T_c$. 
This modification may reduce the temporal oscillation of the low temperature emission measure in the TRAC method observed in Fig. \ref{fig:plvdem_t_typical}.
In this study, we did not use this temporal smoothing to clarify the effect of the different spatial profiles of the broadening factor between the TRAC and LTRAC methods.

In this study, we propose a new method called LTRAC to determine the broadening factor $f_c$ using only the information from nearby grid points.
Similar to the TRAC method, the LTRAC method uses the temperature gradient to detect the unresolved transition region.
We define the normalized temperature gradient as
\begin{equation}
  \label{eq:def_ltrac_r}
  r=\exp\left|\frac{\Delta{s}}{T}\frac{\partial{T}}{\partial{s}}\right|.
\end{equation}
In the unresolved transition region, the quantity $r$ becomes large and the non-negligible numerical artifacts can lead to unphysical solutions (e.g., the wrong density in the corona; see also Fig. \ref{fig:pl0d_typical} and \ref{fig:pl1d_typical}).
To reduce the steep temperature gradient in the high $r$-value region, the broadening factor $f_c$ is determined as
\begin{equation}
  \label{eq:def_fc}
  f_c=\overline{\max(1,r/r_c)^p},
\end{equation}
where $r_c$ and $p$ are non-dimensional parameters, and the overline indicates a low-pass filter in space.
The threshold parameter $r_c{>}1$ is introduced so that the broadening factor $f_c$ is kept unity (no broadening) in the grid points with smooth temperature profile of $r<r_c$.
The parameter $p{\ge}1$ is introduced to accelerate convergence.
We used $r_c=1.5$ and $p=4$ as typical values in this study.
The dependence of the LTRAC method on these parameters is investigated in Appendix \ref{appendix:ltrac_params}.
The low-pass filter is introduced to smooth the spatial distribution of $f_c$ computed from the numerical gradient of the temperature, which is generally non-smooth.
As our discretization described in Sec. \ref{subsec:num_setup} is second-order in space, we used the second-order low-pass filter defined by
\begin{equation}
  \overline{U_j}=\frac{U_{j-1}+2U_j+U_{j+1}}{4},
\end{equation}
where $U_j$ indicates the value of a variable $U$ at the $j$-th grid point along the $s$-coordinate.
We note that the procedure above can be extended easily into multidimensional geometries, which is one of the advantages of the LTRAC method.

Here, we summarize the broadening process of the transition region by the LTRAC method.
When the temperature gradient becomes steeper ($r>r_c$), $f_c$ is enhanced by Eq. (\ref{eq:def_fc}). The enhanced $f_c$ broadens the transition region through the enhanced thermal conduction by Eq. (\ref{eq:limit_kappa_qrad}) while \tred{preserving} the total radiative loss (Eq. (\ref{eq:net_qrad_unchange})). In the broadened transition region, the broadening factor $f_c$ decreases and approaches unity. The value of $f_c$ in the steady state is determined by the balance between the broadening effect by the LTRAC method and the thinning effect by the physical nature of the transition region.

\subsection{Velocity differential emission measure}
\label{subsec:vdem}

To investigate the effect on the spectroscopic observables, we use the velocity differential emission measure (VDEM) for optically thin emissions. The VDEM is defined by
\begin{equation}
  \int_T\left[\int_V\mathrm{VDEM(T,V)}dV\right]dT=\int_hn_en_Hdh,
\end{equation}
where $V$ is the line-of-sight velocity, and the integration on the right-hand-side is taken along the line-of-sight.
Let us focus on the optically thin intensity that can be approximated by
\begin{equation}
  \label{eq:int_coronal_equil}
  I(\lambda)=\int_hC(T,V,\lambda)n_en_Hdh.
\end{equation}
Equation (\ref{eq:int_coronal_equil}) is usually valid if ionization/excitation equilibrium is satisfied in the coronal approximation \citep{2018LRSP...15....5D}.
The VDEM can fully reproduce the intensity, as shown by
\begin{equation}
  I(\lambda)=\int_T\left[\int_VC(T,V,\lambda)\mathrm{VDEM}(T,V)dV\right]dT.
\end{equation}
The effect of the proposed method on the optically thin spectroscopic observables can therefore be analyzed using the VDEM.
The zeroth-, first-, and second-order moments of VDEM with respect to $V$ correspond to the usual differential emission measure (DEM), Doppler velocity, and non-thermal line width, respectively.
The definition of the VDEM given above is identical to that given by \cite{2019ApJ...882...13C}, but it is different from the original version presented by \cite{1995ApJ...447..915N}.

We found that calculations of the optically thin emission that employ the methods for broadening the unresolved transition region require a special care.
As both the TRAC and LTRAC methods use the substitution (\ref{eq:limit_kappa_qrad}), the local radiative cooling rate are reduced by a factor of $f_c$ so that the total radiative loss remains unchanged by the numerical broadening of the transition region (see Eq. (\ref{eq:net_qrad_unchange})).
Similarly, the optically thin emission from the numerical solution must be calculated taking the broadening factor into account:
\begin{equation}
  \label{eq:int_modified_by_fc}
  I(\lambda)=\int_hC(T,V,\lambda)n_en_H\frac{dh}{f_c}.
\end{equation}
The VDEM must be computed in a similar manner.
This correction mainly affects the low temperature region.

\section{Results}
\label{sec:results}

In this section, we analyze the performance of the LTRAC method compared with previous methods.
We investigated three different methods for treating the unresolved transition region: one uses the classical Spitzer-H\"{a}rm conduction without applying any broadening technique (denoted as SH for brevity); one uses the TRAC method; and one uses the LTRAC method.

\subsection{Overall structure}
\label{subsec:overall}

\begin{figure}[ht!]
  \plotone{{\figdir}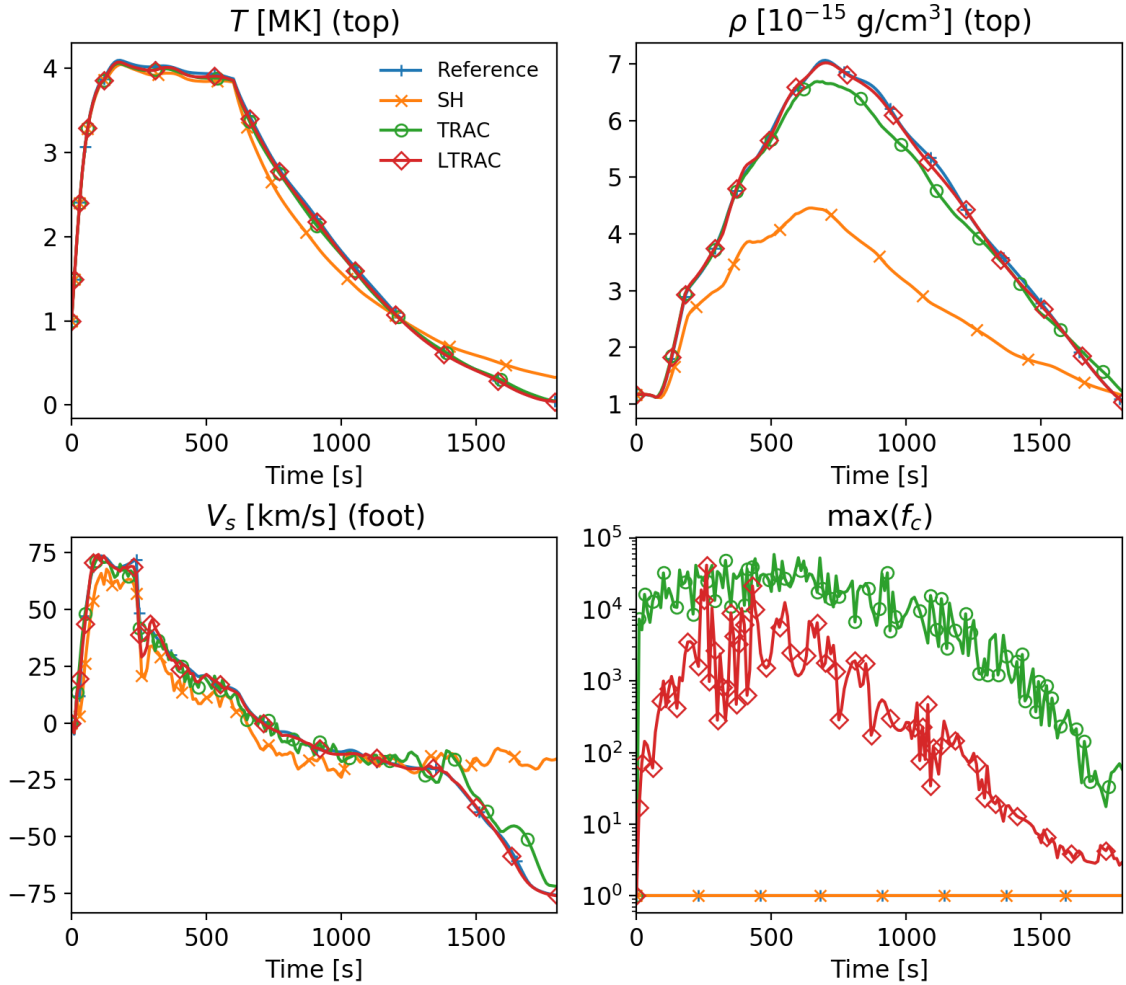}
  \caption{
    Time variation of
    the temperature at the loop top (top left),
    the mass density at the loop top (top right),
    the velocity at the footpoint (bottom left),
    and the maximum value of the transition region broadening factor $f_c$ (bottom right).
    The coronal loop is heated using a spatially uniform volumetric heating term from $t=0$ s to $t=600$ s (see Sec. \ref{subsec:num_setup} for details).
    Shown are the numerical solutions calculated
    using the SH method (orange line with crosses),
    using the TRAC method (green line with circles),
    and using the LTRAC method (red line with diamonds).
    The reference solution is shown as a blue line with plus signs.
    A uniform grid size of $50$ km is used except for the reference.
  }
  \label{fig:pl0d_typical}
\end{figure}

\begin{figure}[ht!]
  \plotone{{\figdir}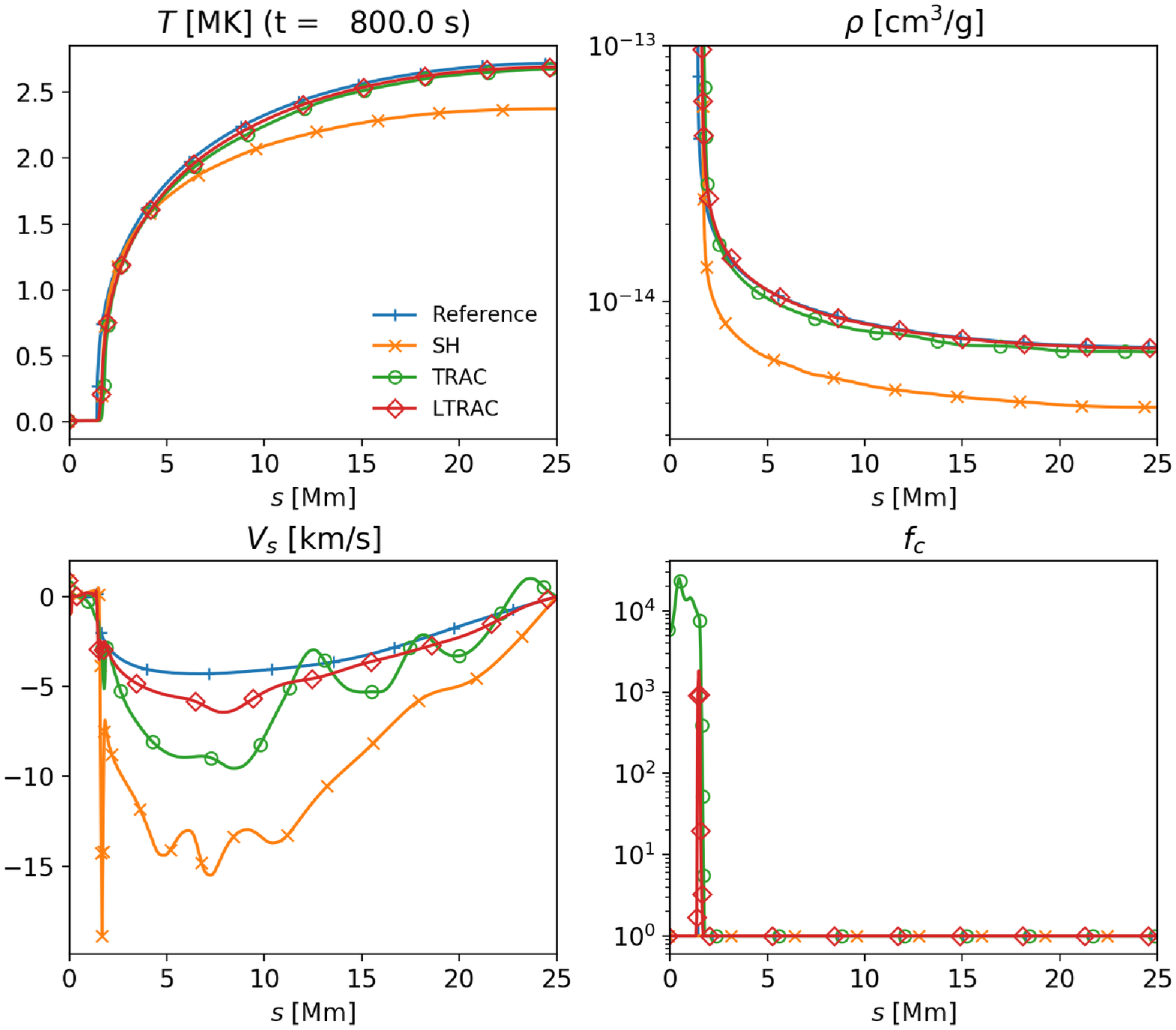}
  \caption{
    Spatial variation along a coronal loop at $t=800$ s.
    Shown are the temperature (top left),
    the mass density (top right),
    the velocity (bottom left),
    and the transition region broadening factor $f_c$ (bottom right),
    The notation is same as that in Fig. \ref{fig:pl0d_typical}.
  }
  \label{fig:pl1d_typical}
\end{figure}

The time variation of the spatially averaged variables calculated with a uniform grid size of $50$ km is shown in Fig. \ref{fig:pl0d_typical}.
The variables at the loop top are averaged over the interval of $0.6L{\le}s{\le}L$, whereas the variables at the footpoint are averaged over $4$ Mm ${\le}s{\le}$ $6$ Mm.
The high-resolution reference solution is calculated using the SH method with a grid size near the transition region of $100$ m.
We refer the reader to Appendix \ref{appendix:reference} for the details of the reference solution.
The temperature of the \tred{under-resolved SH solution (with the grid size of $50$ km)} slightly deviates from the reference solution.
A larger deviation of the \tred{under-resolved} SH solution from the reference solution can be observed in the mass density \citep{2013ApJ...770...12B}.
The mass density at the loop top is more than 30 percent smaller than that of the reference solution.
In contrast, the numerical solutions obtained using the TRAC or LTRAC methods show good agreement with the reference solution.
For the variation of the mass density and velocity, the LTRAC solution is slightly close to the reference solution.
The maximum value of the broadening factor $f_c$ in the LTRAC solution is about one order of magnitude smaller than that in the TRAC solution.
The smaller value of $f_c$ implies smaller thermal conduction coefficients, which is advantageous if the thermal conduction term is integrated explicitly.

\tred{
The maximum value of the broadening factor $f_c$ in the LTRAC method shows temporal oscillation as shown in Figure \ref{fig:pl0d_typical}.
We could not fully understand its origin, but the oscillatory nature may be caused by a kind of feedback between the broadening effect of the transition region by the LTRAC method and the thinning effect from the nature of the transition region. 
For a better understanding, we checked the dependence on the size of time stepping by running the simulation with $10$ times smaller time step size.
If the oscillatory nature of $f_c$ is due to the time discretization error in the evaluation of $r$ and $f_c$ in Eqs. (\ref{eq:def_ltrac_r}) and (\ref{eq:def_fc}), it should depend on the size of time stepping.
However, the temporal variation of $f_c$ in the smaller time step size was almost the same as in Fig. \ref{fig:pl0d_typical}.
%
We also checked the dependence on the LTRAC parameter $p$.
Note that as discussed in Appendix \ref{appendix:ltrac_params}, the larger $p$ is, the larger $f_c$ is expected to be.
If the oscillatory nature of $f_c$ is caused by the feedback between the broadening effect by the thermal conduction (enhanced by $f_c$) and the decrease in $f_c$ (due to the smoothed temperature gradient), the period of the temporal variation of $f_c$ would depend on the parameter $p$.
However, the period of the oscillation did not show strong dependence on the parameter $p$.
Therefore, the oscillatory nature of $f_c$ may not be due to the thermal conduction nor the temporal discretization, but to other processes associated with the thinning of the transition region, like the radiative cooling and/or acoustic waves (related to the hydrostatic equilibrium).
}

Figure \ref{fig:pl1d_typical} shows the spatial variation of the temperature, mass density, velocity, and the transition region broadening factor $f_c$ at the snapshot of $800$ s.
Similar to the spatially averaged variables in Fig. \ref{fig:pl0d_typical}, the \tred{under-resolved} SH solution exhibits the largest difference from the reference solution, and the LTRAC solution is closest to the reference solution.
The spatial profiles of the temperature and mass density obtained using the TRAC and LTRAC methods are close to the reference profile.
In terms of the spatial profile of the velocity, both the TRAC and LTRAC solutions deviate from the reference solution by at least several km.
In addition, the velocity profiles from the TRAC and LTRAC solutions are not as smooth as that of the reference solution, which may be a numerical artifact.
An important difference between the TRAC and LTRAC methods can be observed in the spatial profile of the broadening factor $f_c$.
The value of $f_c$ in the LTRAC solution increases only in the transition region, whereas that value in the TRAC solution increases all the way from the surface ($s=0$) to the transition region.
Considering the good agreement of the LTRAC solution to the reference solution, the increased broadening factor $f_c$ in the upper chromosphere of the TRAC solution is not necessary to reproduce the dynamics and energetics of the coronal plasma.
Moreover, the redundant enhancement of thermal conduction in the upper chromosphere may lead to weak chromospheric evaporation, which causes excess of low temperature emission, as shown in Sec. \ref{subsec:observables}.

\subsection{Effects on spectroscopic observables}
\label{subsec:observables}

\begin{figure}[ht!]
  \plotone{{\figdir}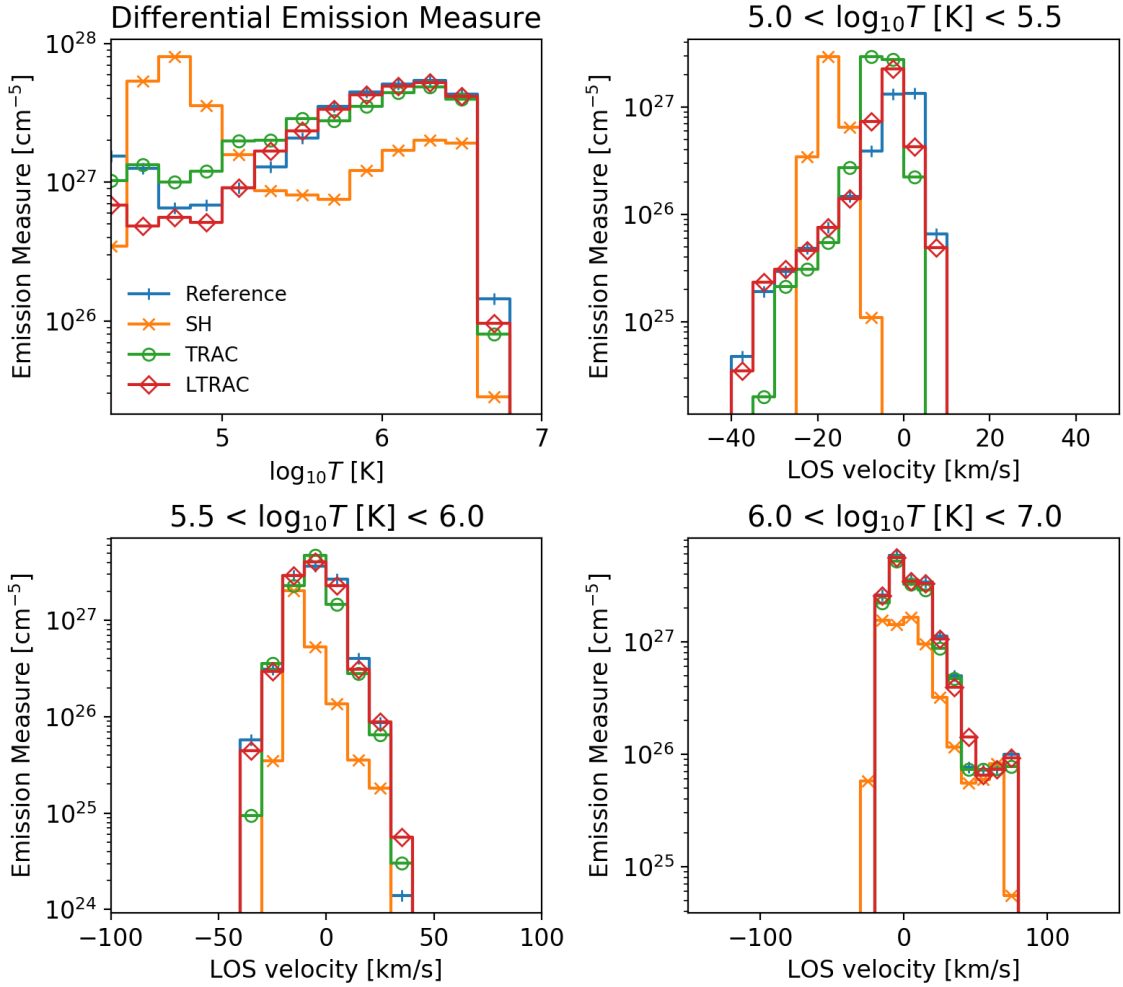}
  \caption{
    Time averaged DEM (top left)
    and the velocity DEM averaged over the temperature intervals of
    $[10^{5.0},10^{5.5}]$ K (top right),
    $[10^{5.5},10^{6.0}]$ K (bottom left),
    and $[10^{6.0},10^{7.0}]$ K (bottom right).
    The notation is same as that in Fig. \ref{fig:pl0d_typical}.
  }
  \label{fig:plvdem_typical}
\end{figure}

\begin{figure}[ht!]
  \plotone{{\figdir}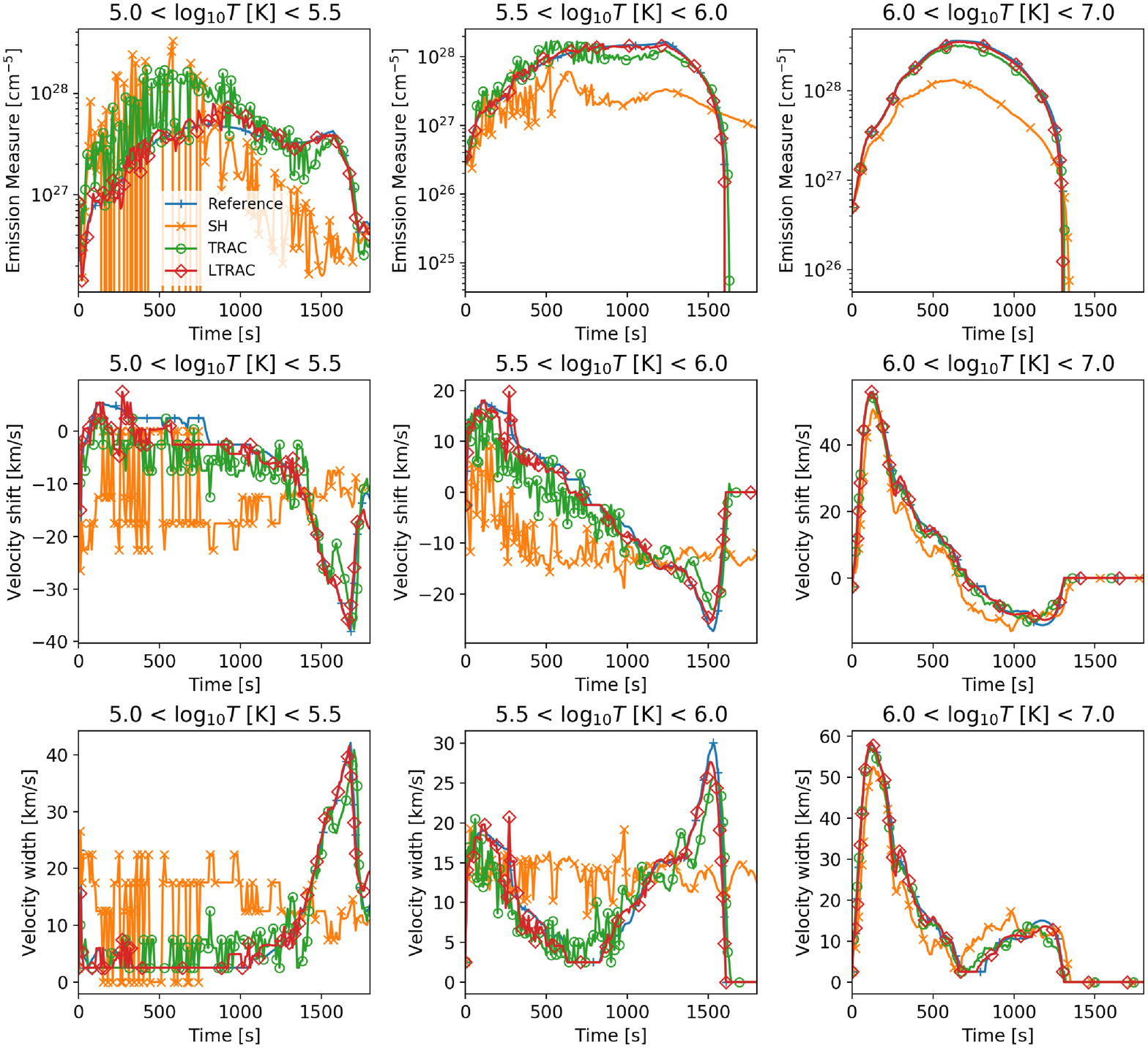}
  \caption{
    Time variation of the emission measure (top row),
    the velocity shift (middle row),
    and the velocity width (bottom row),
    calculated from the VDEM averaged
    over the temperature intervals of
    $[10^{5.0},10^{5.5}]$ K (left column),
    $[10^{5.5},10^{6.0}]$ K (middle column),
    and $[10^{6.0},10^{7.0}]$ K (right column).
    The notation is same as that in Fig. \ref{fig:pl0d_typical}.
  }
  \label{fig:plvdem_t_typical}
\end{figure}

To evaluate the performance of the LTRAC method on spectroscopic observables, we used the velocity DEM described in Sec. \ref{subsec:vdem}.
Figure \ref{fig:plvdem_typical} shows the quantities calculated from the VDEM and temporally averaged from $10$ s to $1500$ s.
The DEM from the \tred{under-resolved} SH solution shows the largest deviation from the reference solution from the transition region temperature to the coronal temperature.
From the resolution dependence within the range of $6.25$ km $\le\Delta{s}\le$ $100$ km, we found that the \tred{under-resolved} SH solution always overestimates (underestimates) the DEM in $T<10^{5.2}$ K ($T>10^{5.2}$ K).
The DEM from the TRAC solution is close to that from the reference solution in $T>10^{5.4}$ K.
However, the TRAC solution overestimates the DEM in lower temperatures of $T<10^{5.4}$ K.
This excess of low temperature emission in the TRAC solution was observed even when a finer grid size was used.
We speculate that enhanced thermal conduction in the upper chromosphere to the lower transition region in the TRAC solution may be heating the high density, cool plasma in the lower layers and enhancing the emission measure in the upper transition region.
In contrast, the LTRAC method successfully reproduces the DEM in the wider temperature range of $T>10^{5.0}$ K.
This result indicates the advantage of the localized broadening factor used in the LTRAC method.
In the lowermost temperature of $T<10^{5.0}$ K, a slight lack of the emission measure can be observed.
The LTRAC solution was found to show smaller DEM even in the higher resolution (down to $\Delta{s}=6.25$ km) in this temperature range.
We also note that the lack/excess of the emission measure in $T<10^{5.0}$ K depends on the parameter $p$ in the LTRAC method (see Appendix \ref{appendix:ltrac_params}).

We averaged the velocity DEM over the temperature intervals $[10^{5.0},10^{5.5}]$, $[10^{5.5},10^{6.0}]$, and $[10^{6.0},10^{7.0}]$ K to mimic a line profile produced by the optically thin emission.
The resulting profile differs from the actual line profile (calculated using the coronal approximation) because the temperature-averaged VDEM profile does not include broadening processes (e.g., thermal or instrumental broadenings).
However, we believe that at least a part of the effect of the LTRAC method on spectroscopic observables can be investigated using the VDEM.
The velocity dependence of the VDEM from the \tred{under-resolved} SH solution deviates from the reference solution, especially in the lower temperature intervals.
The VDEM profile from the TRAC solution agrees well for $T>10^{5.5}$ K with the reference profile, but it deviates slightly in lower temperature.
In contrast, the VDEM profile calculated from the LTRAC solution shows better agreement with that of the reference solution, especially in the low temperature interval.

To evaluate the differences in the VDEM more quantitatively, we calculated the time variation of the velocity moments of the VDEM as shown in Fig. \ref{fig:plvdem_t_typical}.
The velocity shift $V_\mathrm{shift}$ and velocity width $V_\mathrm{width}$ for the temperature interval $T_0<T<T_1$ are defined as
\begin{equation}
  V_\mathrm{shift}=
  \left[\int_{T_0}^{T_1}\int_VV{\cdot}\mathrm{VDEM(T,V)}dVdT\right]
  /\left[\int_{T_0}^{T_1}\int_V\mathrm{VDEM(T,V)}dVdT\right]
\end{equation}
and
\begin{equation}
  V_\mathrm{width}=
  \sqrt{
    \left[\int_{T_0}^{T_1}\int_VV^2\mathrm{VDEM(T,V)}dVdT\right]
    /\left[\int_{T_0}^{T_1}\int_V\mathrm{VDEM(T,V)}dVdT\right]
  },
\end{equation}
respectively.
All of the SH, TRAC, and LTRAC solutions produce oscillatory behavior in the lower temperature interval, because fewer grid points are used for calculating the VDEM in the transition region.
The temporal oscillation of the \tred{under-resolved} SH solution is the most significant because the transition region is resolved by only a few grid points in the numerical solution.
As the TRAC and LTRAC solutions both have more grid points in the transition region, the temporal oscillation is slightly suppressed using these two methods.
Among the three methods, we found that the LTRAC solution exhibits the smallest temporal oscillation and provides the best agreement with the reference solution.
The LTRAC solution can reproduce the velocity shift and width within the difference of several km/s to the reference solution even with a coarse grid size of $50$ km.

\subsection{Convergence analysis}
\label{subsec:converg}

\begin{figure}[ht!]
  \plotone{{\figdir}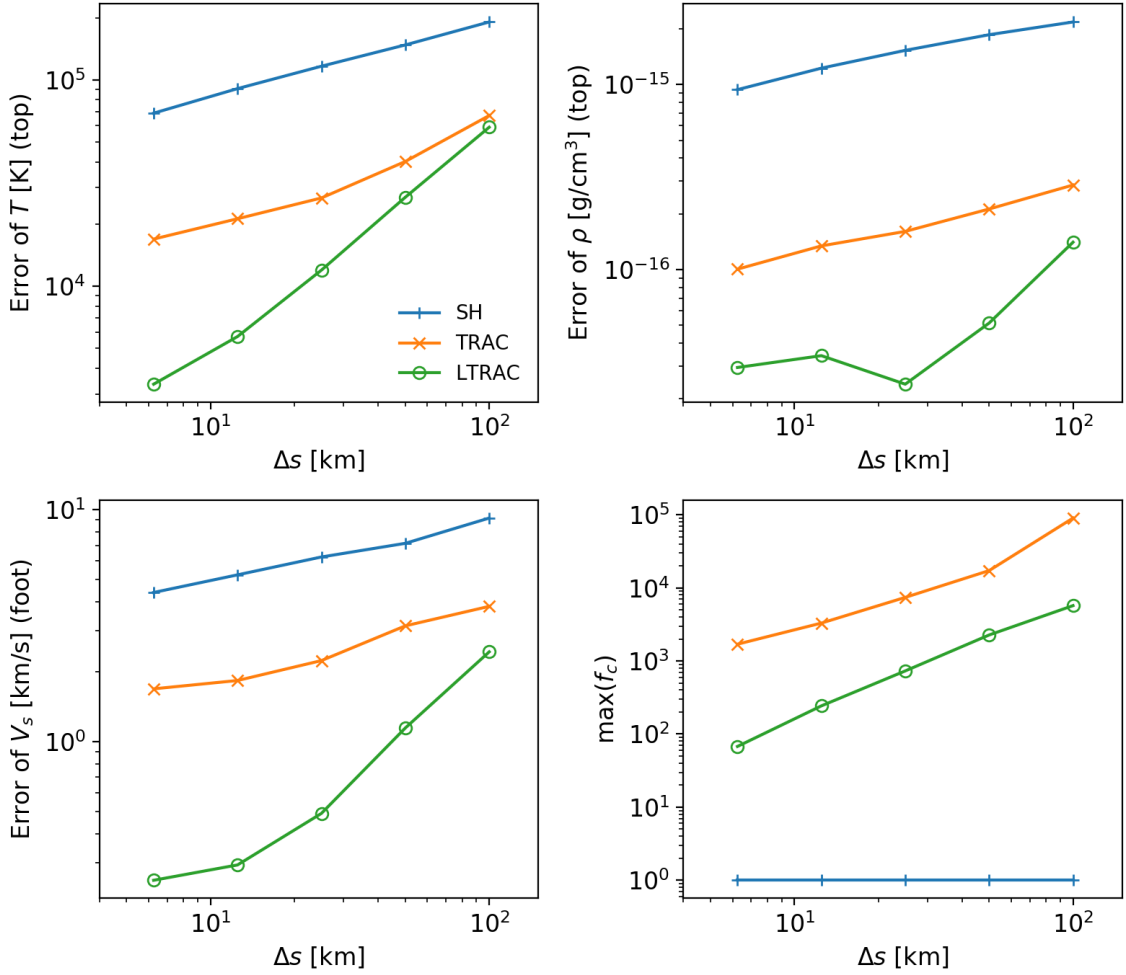}
  \caption{
    Convergence of the spatially averaged variables as function of the grid size.
    Shown are the mean absolute errors measured by
    the temperature at the loop top (top left),
    the mass density at the loop top (top right),
    and the velocity at the footpoint (bottom left),
    as well as the maximum value of the broadening factor $f_c$ (bottom right).
    The results for the SH method (blue line with plus signs),
    the TRAC method (orange line with crosses),
    and the LTRAC method (green line with circles) are shown.
  }
  \label{fig:pl0d_cnvrg}
\end{figure}

\begin{figure}[ht!]
  \plotone{{\figdir}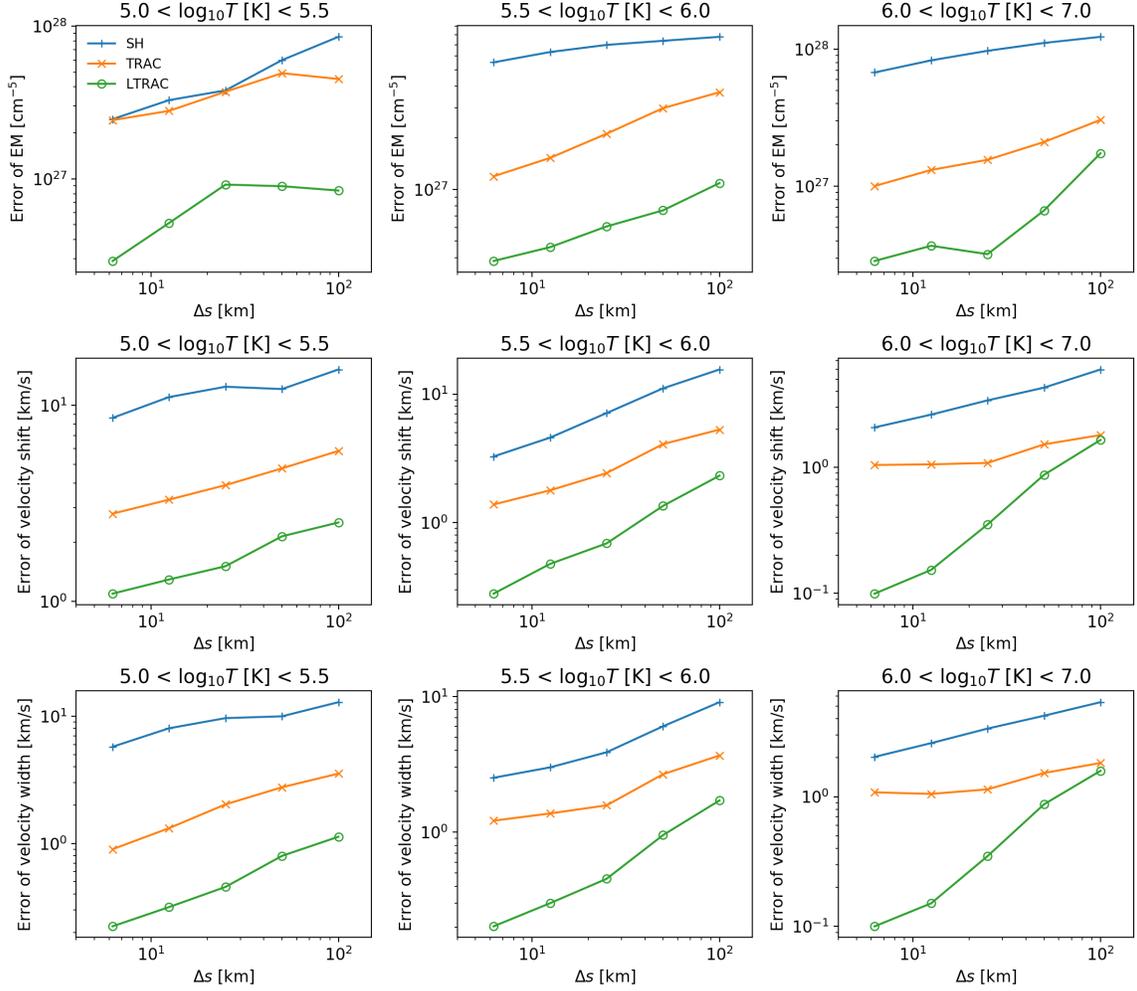}
  \caption{
    Convergence of the VDEM against the grid size.
    Shown are the mean relative error of the emission measure (top row),
    the mean error of the velocity shift (middle row),
    and the mean absolute error of the velocity width (bottom row),
    calculated from the VDEM averaged
    over the temperature intervals of
    $[10^{5.0},10^{5.5}]$ K (left column),
    $[10^{5.5},10^{6.0}]$ K (middle column),
    and $[10^{6.0},10^{7.0}]$ K (right column).
    The notation is same as that in Fig. \ref{fig:pl0d_cnvrg}.
  }
  \label{fig:plvdem_cnvrg}
\end{figure}


Figure \ref{fig:pl0d_cnvrg} shows a convergence analysis of the spatially averaged variables (shown in Fig. \ref{fig:pl0d_typical}).
The error is defined as the mean absolute difference from the reference solution.
The convergence rate of the SH method is very small (less than first-order with respect to the grid size $\Delta{s}$).
The TRAC method shows similarly slow convergence, although the absolute value of the error is much smaller than that of the SH method.
The LTRAC method produces slightly faster convergence of the temperature at the loop top and of the velocity at the footpoint.
The mass density at the loop top in the LTRAC method displays irregular convergence for $\Delta{s}{\le}25$ km.
We speculate that the mass density of the LTRAC simulation with $\Delta{s}{=}25$ km accidentally approached the value in the reference solution.
In the high resolution range of $\Delta{s}{\le}12.5$ km, both the mass density and the velocity of the LTRAC solutions exhibit slow convergence, similar to the SH and TRAC methods.
These results indicate that, although the TRAC and LTRAC methods both yield smaller deviations from the reference solution, the rate of convergence in the high resolution simulation may not be significantly improved from the SH method.
We also measured the dependence of the broadening factor $f_c$ on the grid size by the time average of the spatial maximum.
The values of the broadening factor $f_c$ in the TRAC and LTRAC solutions show approximately a first-order convergence rate.

We also analyzed the convergence of the emission measure, the velocity shift, and the velocity width as shown in Figure \ref{fig:plvdem_cnvrg}.
The LTRAC method shows the smallest error in the emission measure among the three methods, in all three temperature intervals: $[10^{5.0},10^{5.5}]$, $[10^{5.5},10^{6.0}]$, and $[10^{6.0},10^{7.0}]$ K.
The convergence of the emission measure is slow, which may be due to the slow convergence of the mass density shown in Fig. \ref{fig:pl0d_cnvrg}.
The velocity shift and width show better convergence rates (especially in the LTRAC solutions), but the grid convergence are less than first-order.
The errors in the velocity shift and width in the LTRAC solutions are less than $2$ km/s even when a coarse grid size of $50$ km is used.
These results suggest that the Doppler shift and non-thermal line width of the synthesized optically thin line emission from the LTRAC solution may be accurate enough to be compared with spectroscopic observations.


\section{Discussion}
\label{sec:discuss}

From the nature of the transition region, a large number of grid points is required to resolve this thin layer fully.
In this study, we have proposed a new numerical method, called the LTRAC method, which enables the physically accurate coronal dynamics and energetics to be obtained with a coarse grid size.
Following the strategy of \cite{2009ApJ...690..902L} and \cite{2019ApJ...873L..22J}, the LTRAC method broadens the unresolved transition region by modifying the thermal conduction coefficient and the radiative cooling rate.
The major difference between the LTRAC method and the previous methods is that the broadening factor of the transition region is concentrated only in the region with the high temperature gradient.
The localized broadening factor provides better reproduction of the emission measure especially in the lower temperature region, which may result from the suppression of excess conductive heating in the upper chromosphere observed in the previous methods.
We investigated the possible effect of the Doppler velocity and the non-thermal broadening of optically thin emission lines the velocity DEM from the numerical solutions.
We found that the synthesized profile of the optically thin line emission is reproduced within a error of several km/s with a coarse grid size of $50$ km.
The LTRAC method is designed so that it can be extended to multidimensional simulations using the domain decomposition technique.
Application of the LTRAC method to multidimensional coronal heating models \citep[e.g.][]{2011A&A...531A.154G,2015ApJ...812L..30I,2017ApJ...848...38I,2017ApJ...834...10R,2007ApJ...665.1469A} will allow us to reproduce the wave reflection rate at the transition region and the energy loss from the corona more accurately with a coarse grid spacing.

In this paper, we evaluated the performance of the proposed method using one-dimensional hydrodynamic simulations of a coronal loop in an active region.
In flaring loops with higher temperatures, the accuracy of the coarse-resolution simulations may be worse than that for cooler loops.
It should be noted that we assumed a uniform volumetric heating rate in this study.
The performance of the proposed method should be tested with a wider range of numerical settings, including a spatially non-uniform heating rate.
However, we believe that the proposed method will be effective even in such problems, as the free parameters in the LTRAC method determines the sensitivity based only on the normalized temperature gradient and are not chosen for specific problems.

We have ignored the realistic physical processes such as the non-equilibrium ionization or the effect of optical thickness in this study.
The numerically broadened transition region may require the special care on these processes.
For example, in calculating the VDEM (Sec. \ref{subsec:vdem}), we accounted for the broadening factor $f_c$ by reducing the effective line-of-sight distance by a factor of $1/f_c$ (see Eq. (\ref{eq:int_modified_by_fc})), which hopefully conserves the total emission measure from the artificially broadened region.
Similarly, in calculating the optical thickness, the line-of-sight distance must be modified to account for the broadening factor.
Modification non-equilibrium ionization process may be also required.
If we assume that an ion moves across the broadened transition region at a constant velocity, the total number of radiative/collisional reactions the ion undergoes may increase by a factor of $f_c$.
To prevent this artifact, the reaction rates should also be reduced by the broadening factor $f_c$.
We emphasize that these suggestions are mere speculations, and they should be verified more precisely in a future study.

\acknowledgments

H. I. was supported by JSPS KAKENHI grant No. JP19K14756.
This work was supported by the computational joint research program of the Institute for Space-Earth Environmental Research (ISEE), Nagoya University.
The numerical simulations were performed in the Center for Integrated Data Science, Institute for Space-Earth Environmental Research, Nagoya University through the joint research program.

%


\appendix

\section{Convergence of the reference solution}
\label{appendix:reference}

\begin{figure}[ht!]
  \plotone{{\figdir}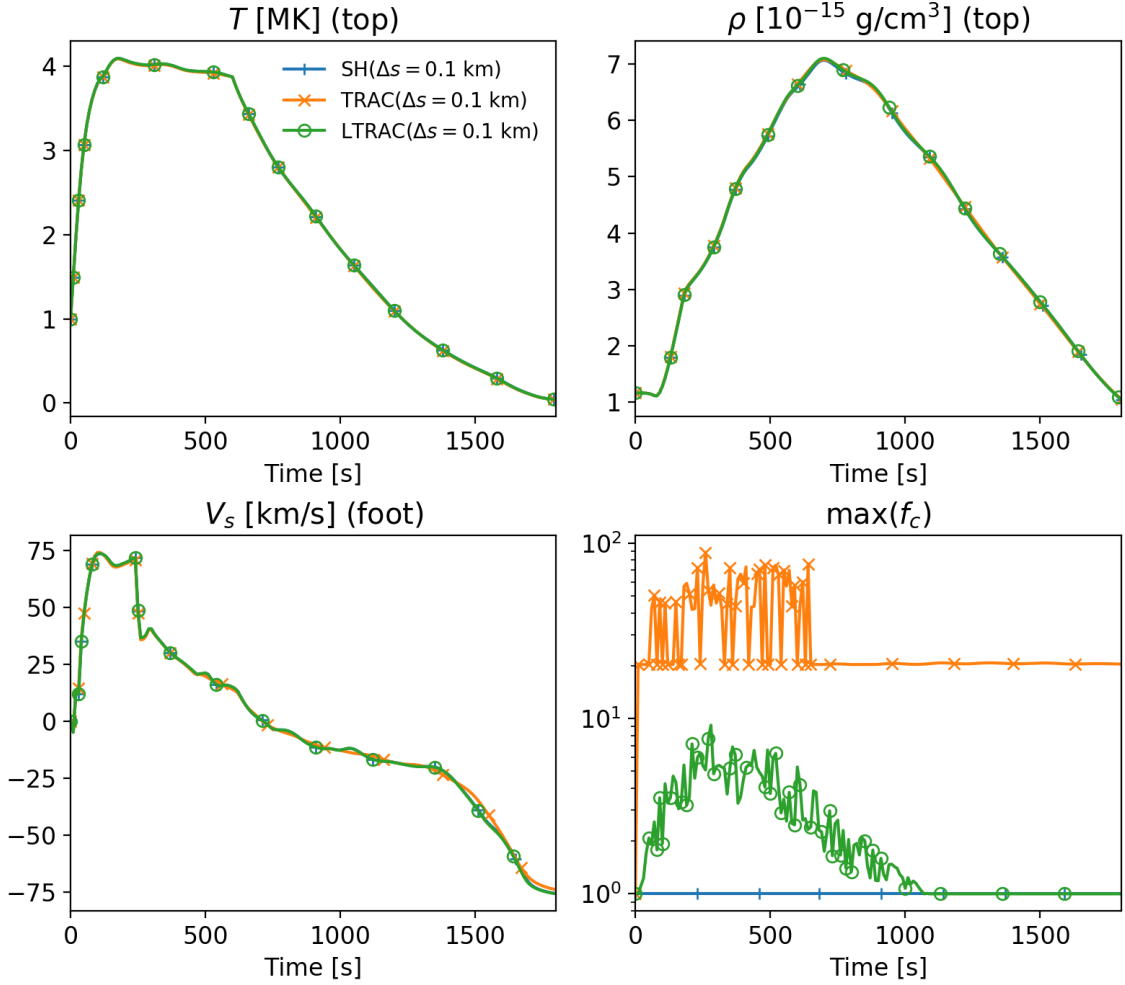}
  \caption{
    Time variation of
    the temperature at the loop top (top left),
    the mass density at the loop top (top right),
    the velocity at the footpoint (bottom left),
    and the maximum value of the broadening factor $f_c$ (bottom right).
    Shown are the numerical solutions calculated
    using the SH method (orange line with times symbol),
    using the TRAC method (green line with circle symbol),
    and using the LTRAC method (red line with diamond symbol).
    The non-uniform grid spacing is used
    with the grid size of $100$ m near the transition region.
  }
  \label{fig:pl0d_01km}
\end{figure}

As the requirement on the grid size is extremely severe in the transition region, we were not able to resolve the transition region fully.
Figure \ref{fig:pl0d_01km} compares high-resolution runs obtained by the SH, TRAC, and LTRAC methods.
The non-uniform grid spacing is used with the grid size of $100$ m near the transition region. 
The variations of the temperature, mass density, and velocity field calculated by these three methods are very close to each other, which implies that the numerical solutions are roughly converged at this grid size.
If the transition region in the numerical solution is fully resolved, the normalized temperature gradient of Eq. (\ref{eq:def_ltrac_r}) should be smaller than the critical value ($\exp(\sigma_c)$ for the TRAC method and $r_c$ for the LTRAC method), which implies that the maximum value of the $f_c$ should be unity in both the TRAC and LTRAC methods.
The nearly constant (but larger than unity) value of $f_c$ in the TRAC solution after $1000$ s occurs because the lower limit of $T_c$ is set to $2\times10^4$ K.
Unfortunately, $\max(f_c)$ in the LTRAC solution is about $5$, which indicates that the numerical grid size should be $20$ m or less to obtain a truly converged solution (in the sense that the LTRAC solution becomes exactly identical to the SH solution).
Such a high resolution simulation is difficult to achieve even in a one-dimensional domain as we do not use the adaptive mesh refinement.
This requirement that the grid size be less than $20$ m is consistent with the estimate obtained using the static loop model of \cite{1978ApJ...220..643R}, which predicts that the thickness of the transition region is about $400$ m for a loop top temperature of $4$ MK (see Sec. \ref{sec:intro}).
The results presented in this paper do not depend on the choice of the reference solution, except for the the convergence rates of the highest resolution runs (grid sizes less than $20$ km) in Appendix \ref{appendix:ltrac_params}.
In this study, we use the SH solution with the grid size of $100$ m as the reference solution.

\section{Dependence on LTRAC parameters}
\label{appendix:ltrac_params}

\begin{figure}[ht!]
  \plotone{{\figdir}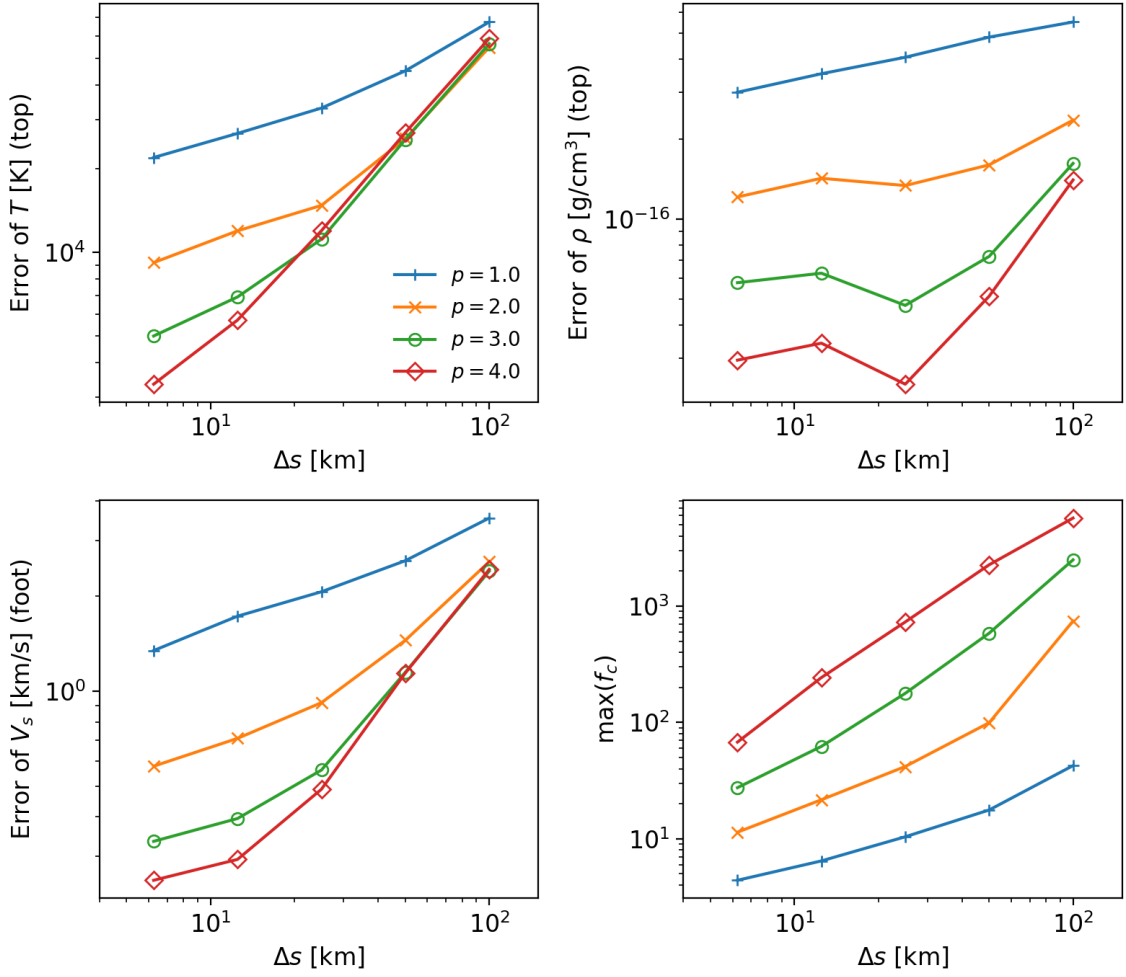}
  \caption{
    Same as Fig. \ref{fig:pl0d_cnvrg} but showing
    the dependence on the LTRAC parameter $p$.
    The results for $p=1$ (blue line with plus signs)
    $p=2$ (orange line with crosses),
    $p=3$ (green line with circles),
    and $p=4$ (red line with diamonds) are shown.
  }
  \label{fig:pl0d_pdep}
\end{figure}

\begin{figure}[ht!]
  \plotone{{\figdir}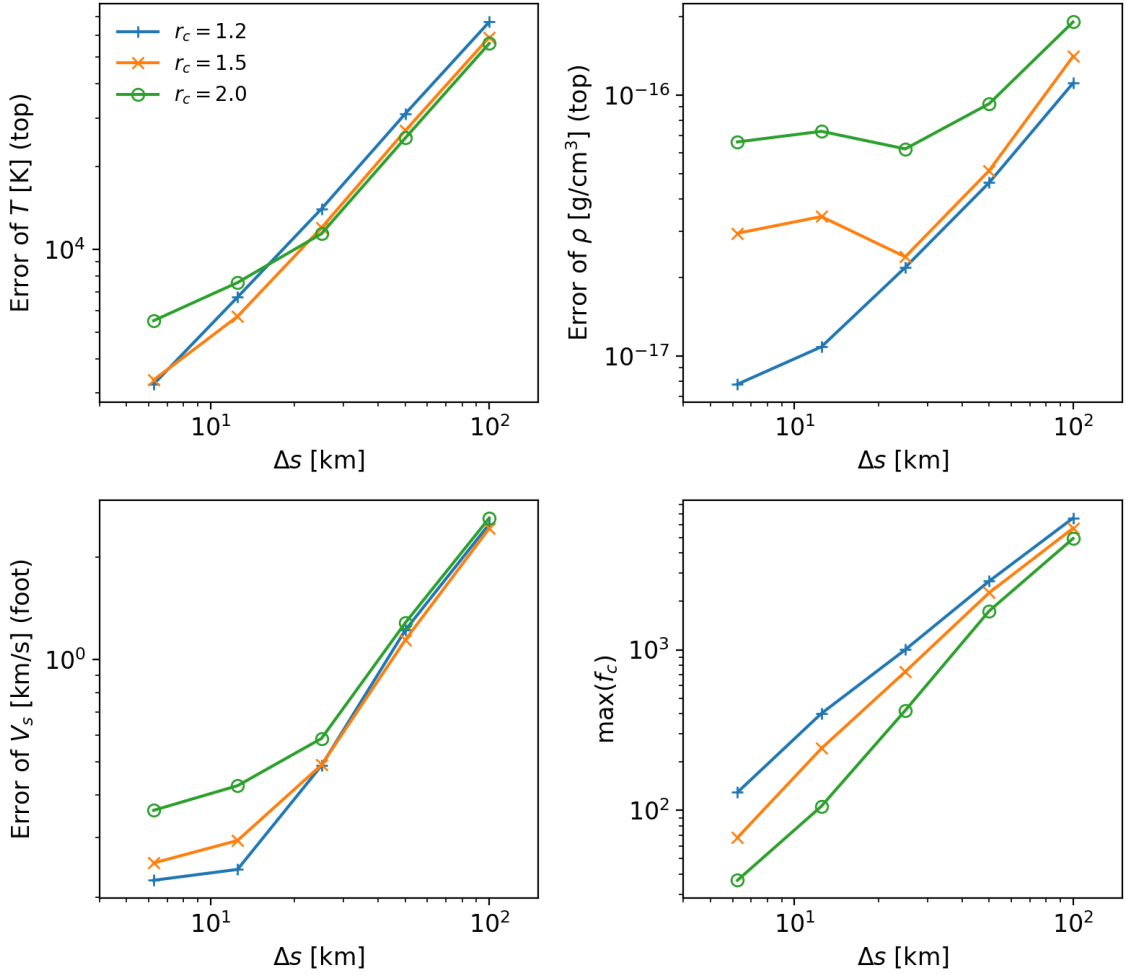}
  \caption{
    Same as Fig. \ref{fig:pl0d_cnvrg} but showing
    the dependence on the LTRAC parameter $r_c$.
    The results for $r_c=1.2$ (blue line with plus signs)
    $r_c=1.5$ (orange line with crosses),
    and $r_c=2$ (green line with circles) are shown.
  }
  \label{fig:pl0d_rdep}
\end{figure}

\begin{figure}[ht!]
  \epsscale{0.7}
  \plotone{{\figdir}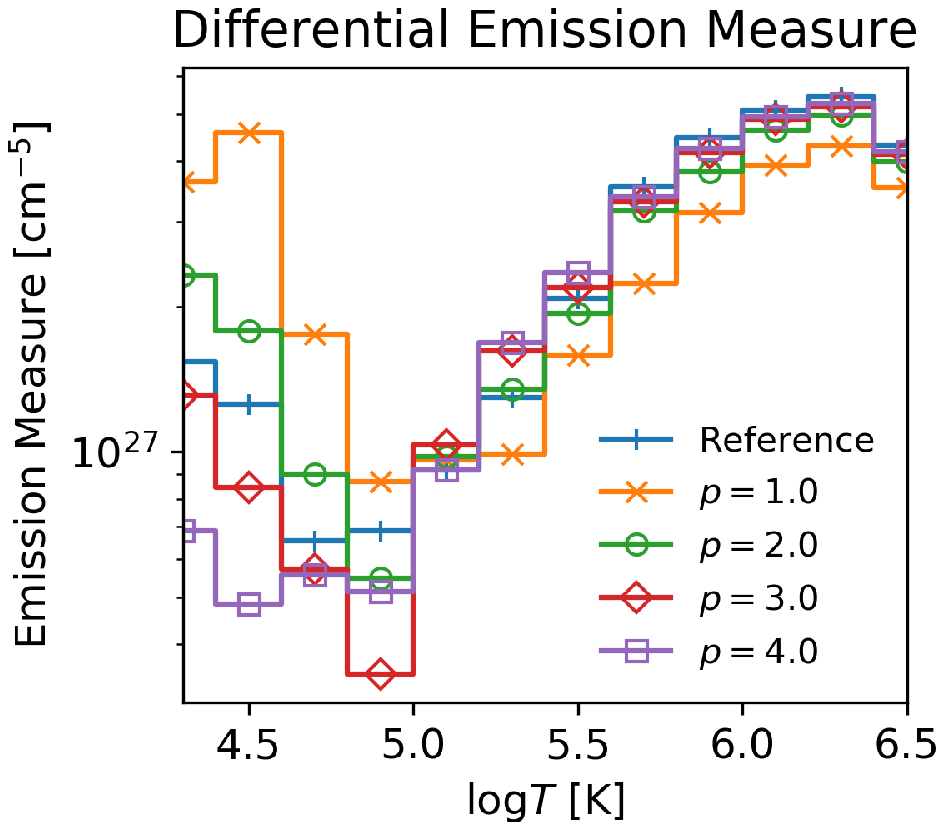}
  \caption{
    Dependence of the time averaged DEM on the LTRAC parameter $p$.
    The results for $p=1$ (orange line with crosses),
    $p=2$ (green line with circles),
    $p=3$ (red line with diamonds),
    $p=4$ (purple line with boxes),
    and the reference (blue line with plus signs) are shown.
  }
  \label{fig:pldem_pdep}
\end{figure}



The LTRAC method employs two parameters, the critical value $r_c$ of the temperature gradient and the acceleration factor $p$.
Here, we briefly investigate the dependence of the LTRAC solutions on these parameters.
Figure \ref{fig:pl0d_pdep} shows the dependence of the convergence rates on the LTRAC parameter $p$, fixing the value of $r_c$ to be $1.5$.
Clearly, a larger value of $p$ produces faster grid convergence.
The error in the LTRAC method with $p=2$ is similar to that of the TRAC method (Fig. \ref{fig:pl0d_cnvrg}).

Figure \ref{fig:pl0d_rdep} shows the dependence of the convergence rate on $r_c$, with the value of $p$ fixed to $4.0$.
The result is not very sensitive to the parameter $r_c$, although a smaller $r_c$ tends to produce smaller errors and larger values of the broadening factor $f_c$.
The reader may notice that the mass density of $r_c=1.2$ apparently converges to the reference value regularly.
However, we note that this reqular convergence may be produced by the insufficient resolution of the reference solution.
If we use the high-resolution LTRAC run as a reference, the mass density shows irreqular convergence in all values of $r_c$.
We chose the typical value $r_c=1.5$ so that the threshold of the temperature gradient would be close to the value in the TRAC method (so as to $\exp(\sigma_c){\sim}r_c$).

In Sec. \ref{subsec:vdem}, we have mentioned a small lack of the low temperature DEM in $T<10^{5.0}$ K found in the LTRAC solution.
The lack and excess of the DEM in this low temperature range depends on the parameter $p$.
Fig. \ref{fig:pldem_pdep} shows the parameter dependence of the DEM.
The uniform grid size of $50$ km is used except the reference solution.
In the cases of smaller $p$-value (i.e., $p{\le}2$), the emission measure in $T<10^{5.0}$ K tends to show slight excess, in contrast to the cases for $p{\ge}3$.
This dependence on the parameter $p$ was observed independently of the spatial resolution (down to $\Delta{s}=6.25$ km).
This result may imply that the optimal parameter of $p$ lies between $2$ and $3$ in terms of the DEM reproducibility.
We found that the dependence of the low temperature DEM on the parameter $r_c$ is not significant.



\end{document}